\documentclass{elsart3}

\usepackage{graphicx}

\usepackage{amssymb}

\begin{document}

\begin{frontmatter}



\title{Design and Performance of the CMS Pixel Detector Readout Chip}


\author[psi]{H.Chr. K\"astli\corauthref{cor}},
\author[uniba,psi]{M. Barbero},
\author[psi]{W. Erdmann},
\author[unizh,psi]{Ch. H\"ormann},
\author[psi]{R. Horisberger},
\author[psi]{D. Kotlinski},
\author[eth]{B. Meier}

\address[psi]{Paul Scherrer Institut, 5232 Villigen PSI, Switzerland}
\address[uniba]{Institut f\"ur Physik der Universit\"at Basel, 4056 Basel, Switzerland}
\address[eth]{Institut f\"ur Teilchenphysik, ETH Z\"urich, 8093 Z\"urich, Switzerland}
\address[unizh]{Physik-Institut der Univerit\"at Z\"urich, 8057 Z\"urich, Switzerland}
\corauth[cor]{Corresponding author.\\ {\em Email: hans-christian.kaestli@psi.ch}}

\begin{abstract}
The readout chip for the CMS pixel detector has to deal with an enormous
data rate. On-chip zero suppression is inevitable and hit data must be
buffered locally during the latency of the first level trigger. Dead-time must be kept at a minimum. It is dominated 
by contributions coming from the readout. To keep it low an analog readout scheme has been adopted where pixel addresses are analog coded.\\
We present the architecture of the final CMS pixel detector readout chip with special emphasis on the analog readout chain. Measurements of its performance are discussed.
\end{abstract}

\begin{keyword}
CMS \sep Pixel Detector \sep Readout Chip
\PACS 29.40.Wk \sep 29.40.Gx \sep 85.40.-e
\end{keyword}
\end{frontmatter}

\section{Introduction}
CMS is a general purpose detector at the Large Hadron Collider (LHC) at CERN. Its innermost tracking device is the pixel detector, which consists of three barrel layers and two endcap disks both up and downstream\cite{TDR}. Its basic building blocks are highly segmented silicon sensors with corresponding readout chips (ROCs).  Each sensor segment (pixel cell) is connected to a charge sensitive amplifier in the corresponding pixel unit cell (PUC) of the ROC via an indium bump of about $20\mu$m diameter.
In the first barrel layer, the active area is 4.3cm from the interaction point. At this short distance the density of produced particles will be as high as $4\cdot 10^7$sec$^{-1}$cm$^{-2}$ at the full LHC design luminosity of $\mathcal{L}=10^{34}$sec$^{-1}$cm$^{-2}$. Due to inclined tracks and Lorentz drift of the produced signal charge in the 4T magnetic field several pixels might be hit by a single track forming hit clusters. In order to improve position reconstruction for clusters through signal interpolation the amount of charge produced in each pixel is measured. The ROCs have to record position and charge for all hit pixels with a time resolution of 25ns, which is the time between two LHC bunch crossings. These information have to bo stored on-chip during the CMS first level trigger latency of 3.2$\mu$s after which they are either read out or discarded. The ROCs are read out serially via 40MHz analog links.\\
The first prototype of the readout chip has been developed in the radiation hard DMILL process \cite{DMILL,kurt}. In order to improve performance and yield and to reduce costs, the design has been migrated to a commercial process with much smaller feature size\cite{Wolfram}. The second iteration of the translated chip, PSI46V2, is going to be the production version and is described in this article. \\
In section two the architecture of PSI46V2 is described. Section three focusses on the analog readout scheme and in section four measurements of the performance are presented.

\section{Chip architecture}

The readout chip PSI46V2 has been fabricated in a commercial 5-metal layer 0.25$\mu$m process available through CERN. The design has been made radiation tolerant by following special layout rules as proposed in \cite{radhard}. The chip integrates 1.3 million transistors in an area of 7.9mm$\times$ 9.8mm.  It can be divided into 3 functional blocks: a control and supply block in the chip periphery, an array of pixel unit cells organized in double columns, and the double column peripheries which control readout and trigger validation within double columns. The total number of pixels is 80$\times$52 with a pixel size of $100\mu$m $\times 150 \mu$m.

\subsection{Control and supply block}
The chip periphery houses various control and supply circuits. It contains
\begin{itemize}
\item A serial programming interface. This is an I$^2$C-like protocol, modified to run at a speed of 40 MHz. To accommodate this high speed the possibility to read back configuration data had to be given up. Nevertheless, there is a limited readback possibility through the analog data stream as will be shown below. Two low voltage differential signal (LVDS) pairs are needed for clock and data lines.
\item A fast signal decoder. First level triggers and commands for reset and calibration signal injection are coded into a single LVDS signal. This signal has to be decoded and distributed over the chip. 
\item 21 8-bit digital-to-analog converters (DACs), five 4-bit DACs and one 3-bit DAC to adjust offsets, gains, thresholds, supply voltages, timings, etc.
\item 2 control registers to set the trigger latency, readout speed (40 MHz/20MHz) and range for calibration pulses and to enable/disable the ROC.
\item A bandgap voltage reference and 6 voltage regulators. Three of them can be wire-bonded to external filter capacitors. This leads to a good immunity against power ripple and reduces chip-to-chip cross talk. The ROC needs two external power supplies, 1.5V for the analog section and 2.5V for digital part. It consumes a total of about 120 mW which corresponds to only $29\mu$W per pixel. The voltage regulators are programmable and hence the voltages can be set for each chip separately. 
\item An analog event generator. This is a circuit that collects the pixel hit information from the double columns and generates the output data stream as described in section 3.
\item An on-chip temperature sensor for monitoring cooling performance.
\item A cluster multiplicity counter. This is a fast trigger signal that could be used by the CMS first level trigger or for self-triggering in the lab when no external trigger is available. Two thresholds can be set for this mechanism: one sets the minimal number of hits within a double column below which hits are considered as background and are ignored, whereas the other one tunes the number of hit double columns above which a trigger signal is issued. Note, that even below this threshold a signal proportional to the number of clusters is available. 
\end{itemize}

\subsection{Pixel unit cell}
\begin{figure*}
\begin{center}
\includegraphics[angle=-90,width=15cm]{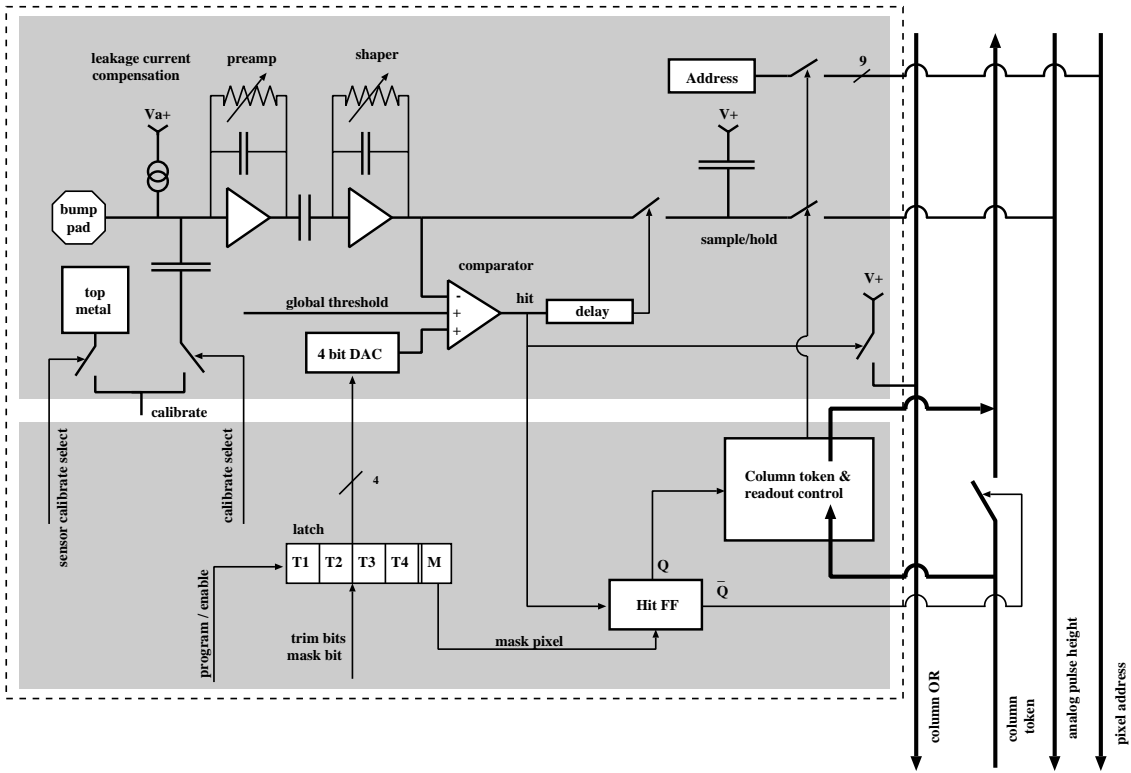}
\end{center}
\caption{Schematic view of a pixel unit cell}
\label{fig:puc}
\end{figure*}
\begin{figure}
\begin{center}
\includegraphics[width=6.5cm]{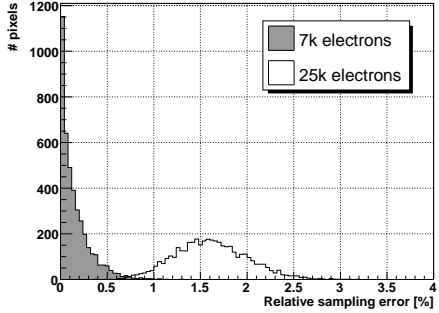}
\end{center}
\caption{Relative error of peak detection across a full chip for 2 different pulse heights}
\label{fig:error}
\end{figure}

A sketch of the PUC is shown in figure \ref{fig:puc}. It can be divided into an analog part (top) and the digital logic (bottom). The thick lines on the right are bus lines running along the double column.\\
The signal from the sensor enters a two stage charge sensitive pre-amplifier/shaper system. Alternatively, calibration signals can be injected through either a 4.8fF injection capacitor connected directly to the amplifier input node, or via the sensor through the air gap between a top metal plate in the ROC and the sensor. The former is used for trimming the comparator threshold while the later can be used to test the bump bond connections\cite{andrei}. A globally programmable current source at the input node compensates for sensor leakage current. \\
Zero suppression is performed with a comparator. A global threshold can be programmed for the whole chip. In order to compensate for local transistor mismatches each pixel has a 4-bit DAC to trim the threshold. Furthermore, a mask bit allows to disable noisy pixels. Once the comparator is above threshold the shaper output signal is stored in a sample-and-hold circuit (see below). The double column periphery is notified immediately through a fast hard-wired column OR. The pixel becomes insensitive and waits for a column readout token. When it arrives the analog signal is sent to the periphery together with the pixel row address. The token is then passed on and the pixel resumes data taking. Thus, dead-time is short but depends on the hit rate. The time needed for a column scan to finish is $\leq$ 50ns + (50ns$\times$ number of hits). \\
The shaper output signal is sampled after a fixed time delay which is controlled by a global register (i.e. once per ROC). The signal within the sample-and-hold circuit follows the rising edge of the shaper output quite fast while it decays much slower. This makes the peak of the signal sufficiently flat and the circuit is not very sensitive to this time delay. Due to timewalk and device mismatch the exact peaking time depends somewhat on the pulse height and varies a bit from pixel to pixel. The relative error in the measured pulse height due to a non perfect sampling point has been measured and is shown in figure \ref{fig:error}. If the delay is optimised for low signal charges (3500 electrons in figure \ref{fig:error}), the relative error for large signals is only a few per cent.

\begin{figure}[b]
\begin{center}
\includegraphics[width=4cm]{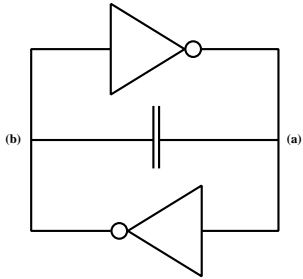}
\end{center}
\caption{SEU protected storage cell}
\label{fig:SEU}
\end{figure}

\noindent Downloading the full detector configuration at run start will take $\lesssim$ 0.5 minute \cite{danek}. If needed, pixels can be reprogrammed during LHC machine orbits reserved for sub-detector calibration, however only at a limited rate. Thus single event upset (SEU) in the trim and mask bit storage cells has to be considered. A simple trick has been adopted to protect storage cells from SEU. This is illustrated in figure \ref{fig:SEU}. 
If node (a) collects a large enough amount of charge, produced by a nuclear reaction, the stored logic level can be changed. The presence of the capacitor aided by Miller effect increases the critical charge and thus reduces the probability of an upset.

\noindent The SEU cross section for both protected and unprotected storage cells has been measured at PSI for 300 MeV/c pions. The result is $\sigma=2.6\cdot 10^{-16}$cm$^2$ per storage cell for protected and $\sigma=2.4\cdot10^{-14}$cm$^2$ for unprotected cells. This is an improvement of 2 orders of magnitude. In PSI46V2 all storage cells in the PUCs are protected (4 trim bits and one mask bit). In case of an SEU the flipped bits have to be reprogrammed through control links. Each link serves several modules/pannels. In the worst case one link controls 4 modules plus 4 half-modules in each of the two innermost barrel layers, i.e. a total of 192 ROCs or 96 ROCs per layer. At high luminosity LHC operation the expected SEU rate for this control link is $<3\cdot 10^{-2}$Hz. This means that in about 8 hours 0.1\% of the pixels changed their threshold. The occupancy of each pixel can be monitored online and pixels that show significant changes will be reprogrammed.

\subsection{Double column periphery}
\begin{figure*}[t]
\begin{center}
\includegraphics[width=16cm]{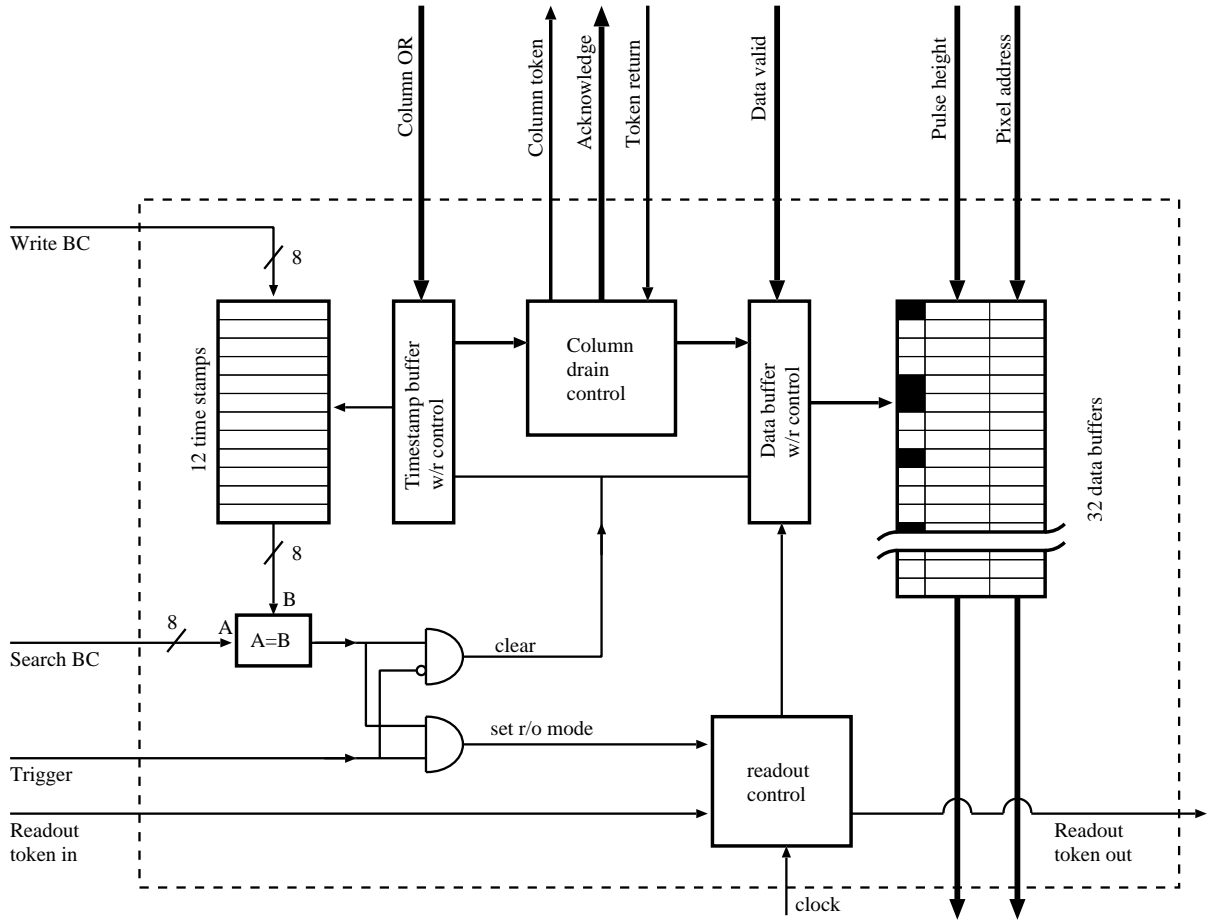}
\end{center}
\caption{Schematic view of the double column periphery}
\label{fig:peri}
\end{figure*}

The double column periphery controls the transfer of hit information from the pixels to the storage buffers (column drain mechanism) and performs trigger verification. A schematic view of the logic is shown in figure \ref{fig:peri}.\\ 
When the asynchronous fast column OR signal arrives at the periphery three actions are performed:
\begin{enumerate}
\item[1. ] The present value of a bunch crossing counter (WBC) is latched into one of 12 time stamp buffer cells. This is needed later on for trigger validation.
\item[2. ] The hits for this bunch crossing are acknowledged by the periphery. This tells the pixels to associate any later hits with another column drain and its corresponding bunch crossing. There are one active and up to two pending column drains allowed. Hits of any further bunch crossing are lost.
\item[3. ] A column drain is initiated by sending out a readout token to the first pixel.\\
\end{enumerate} 
A pixel without a hit belonging to the active column drain just bypasses the token. This has been measured to happen at a rate of about 3.3 GHz, i.e. the token arrives at the first hit pixel within less than 50ns or 2 clock cycles. Hit information is transferred in parallel into the data buffer, one pixel per two clock cycles. The data buffer has a depth of 32 units. Each unit consists of a marker bit one analog and nine digital storage cells for pulse height and row address respectively. The marker bit indicates the beginning of a new event and thus is used to synchronize entries in the time stamp and the data buffers. \\
The oldest entry in the time stamp buffer is permanently compared to a second bunch crossing counter (SBC) which is delayed with respect to the WBC by a programmable amount which corresponds to the first level trigger latency. In case of agreement the time stamp is deleted and the presence of the CMS first level trigger signal is checked. Either the corresponding data is discarded or the column is set into readout mode. In this mode, the double column stops data acquisition in order to prevent overwriting of valid data. It waits for the readout token to arrive and sends data to the chip periphery. Afterwards the double column resets itself and hence cannot have further valid hits for the duration of the trigger latency. Data loss also occurs when one of the buffers is completely filled up. In case of a full data buffer a double column gets reset. If the time stamp buffer is full, data acquisition is paused until the next buffer cell is freed.

\section{Readout scheme and chip readout format}
\begin{figure}
\begin{center}
\includegraphics[width=7.5cm]{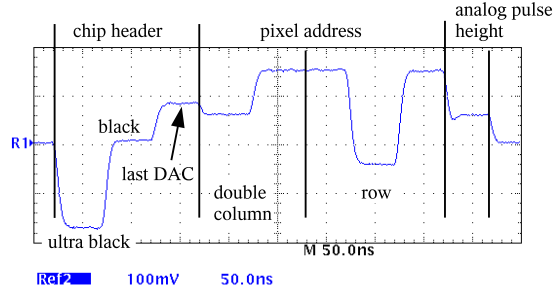}
\end{center}
\caption{Readout sequence of a ROC with one pixel hit}
\label{fig:ro}
\end{figure}

\begin{figure}
\begin{center}
\includegraphics[width=6cm]{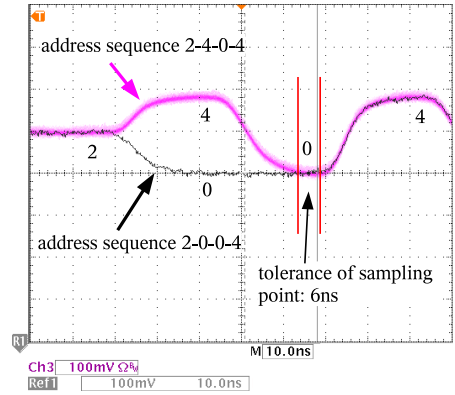}
\end{center}
\caption{Readout signal output of a barrel module for two different address sequences} 
\label{fig:memory}
\end{figure}

\begin{figure}
\begin{center}
\includegraphics[width=7.5cm]{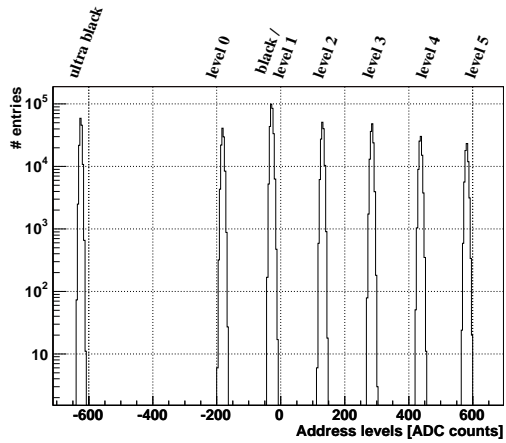}
\end{center}
\caption{Histogram of the sampled address levels of a ROC. The levels are cleanly separated} 
\label{fig:levels}
\end{figure}

\begin{table}[b]
\small{
	\begin{tabular}{|l|c|c|c|}\hline
	Radius [cm] & \makebox[1cm]{4} & \makebox[1.2cm]{7} & \makebox[1.2cm]{11}\\ \hline
	pixel busy 	& 		0.21\% & 	0.078\% & 	0.044\% \\
	column drain busy & 		0.25\% & 	0.020\% & 	0.004\% \\
	time stamp buffer overflow & 	0.17\% & 	0.001\% & 	0 \\
	data buffer overflow & 		0.17\% & 	0.081\% & 	0.065\% \\ \hline
	1-buffer	&		1.0\% &		0.40\% &	0.26\% \\
	readout 	&		1.0\% &		0.20\%	&	0.16\%	\\		
	double column reset & 		1.0\%  &	0.44\% &	0.26\% \\ \hline
	total &				3.8\% &		1.2\% &		0.79\% \\ \hline
	\end{tabular}
}
\label{table:loss}
\caption{Summary of simulated data loss for barrel modules at a luminosity of $10^{34}$sec$^{-1}$cm$^{-2}$ and 100kHz trigger rate. The upper part shows column drain losses, in the lower part readout related data losses are summarized}
\end{table}
A couple of design inherent data loss mechanisms have been mentioned so far. They are summarized in table \ref{table:loss} for barrel modules, a luminosity of $10^{34}\cdot$sec$^{-1}$cm$^{-2}$ and $100$ kHz first level trigger rate. A large contribution, labeled 'readout', comes from the fact that a double column is insensitive between the time it validates a trigger and the time it finishes the readout. This stresses the importance of having short readout times. Therefore an analog readout scheme running at 40 MHz (alternatively it can be switched to 20 MHz) has been adopted where the pixel address is coded into 6 discrete levels. Pixel address and signal pulse height can be transferred in 6 clock cycles. \\
In the endcap part of the detector several sensor tiles of different size are organized in so called blades which contain 24 ROCs on the front side and 21 ROCs on the back side. In the barrel, sensors connected to 16 (8) ROCs are called modules (half modules). The ROCs of each side of a blade or of a (half) module are read out in a daisy chained way. A token bit, controlled by a special token bit manager chip (TBM,\cite{ed}), is sent to the first ROC. The ROC sends its data to the TBM and passes the token bit to the next ROC, until the token bit gets back to the TBM. The TBM amplifies the signals from the ROCs, adds a header and a trailer and drives the signals out to the detector supply tube. \\
The readout of a single pixel hit is shown in figure \ref{fig:ro}. It starts with a header of 3 clock cycles. A large negative signal level well outside of the range of pixel data (ultra-black) followed by a zero differential level (black) separates the individual ROCs in the blade/module data stream. No other identification is sent. In order to get the ROC ID the DAQ unit has to count the chip headers. In the third clock cycle of the chip header a signal inversely proportional to the value of the last addressed DAC is sent. This is the only way information about the configuration of the ROCs can be read back. It follows the double column address (2 cycles), the row address (3 cycles) and the analog pulse height for each hit pixel.
Figure \ref{fig:memory} shows the signal output of a barrel module as it arrives at the electro-optical converter. Shown are two different address level sequences. As can be seen, the risetime of the signals is short enough to allow for a tolerance of the sampling point in the DAQ unit of $\approx$ 6ns. The measured sampling distribution for a ROC is shown in figure \ref{fig:levels}. The address levels are clearly separated and allow a reliable reconstruction of the pixel addresses.

\begin{figure}
\begin{center}
\includegraphics[width=7.5cm]{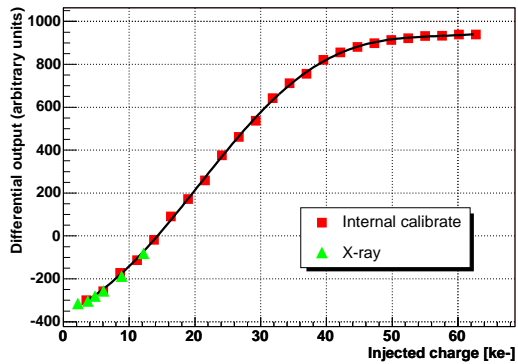}
\end{center}
\caption{Analog signal transmission.}
\label{fig:anout}
\end{figure}
\begin{figure*}
\begin{center}
\includegraphics[width=16cm]{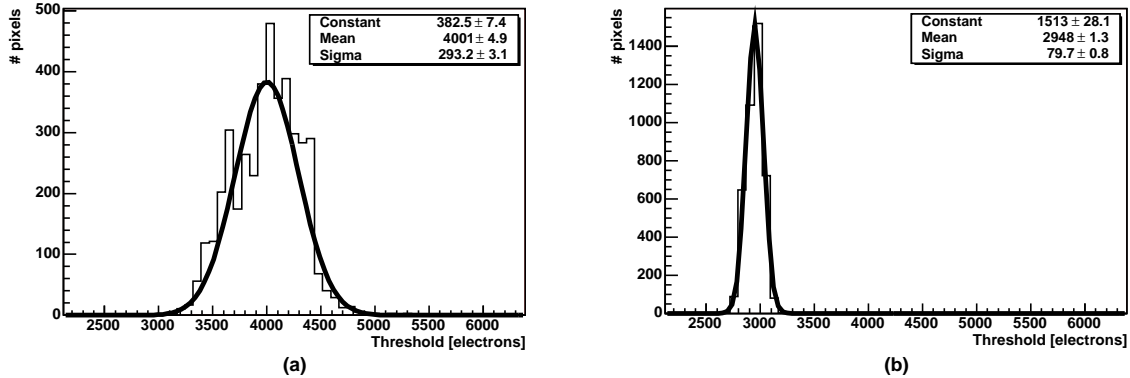}
\end{center}
\caption{Threshold distribution of a ROC before (a) and after (b) trimming}
\label{fig:trim}
\end{figure*}

\section{Measurements}
\begin{figure}[h]
\begin{center}
\includegraphics[width=7cm]{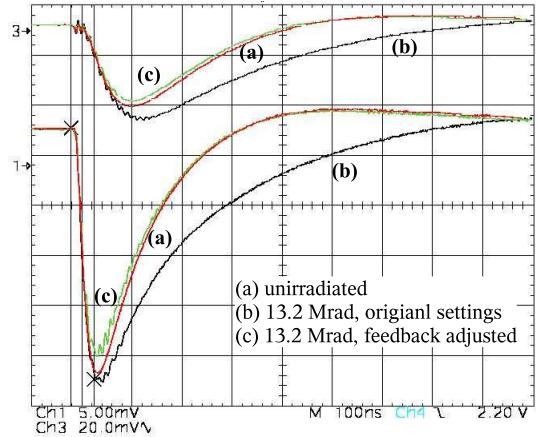}
\end{center}
\caption{Output of sample-and-hold circuit (top) and shaper output (bottom) before and after irradiation.}
\label{fig:irrad}
\end{figure}

In this section lab measurements of the performance of the ROC are shown. 
Figure \ref{fig:anout} shows the pulse height response to calibration signal injection. It is slightly non-linear at low charges and saturates for high charges above about 2 MIPS. 
Injection capacitances have been calibrated with X-ray sources. K$_\alpha$ lines of 6 different isotopes between 8.04keV and 44.2keV have been used. These measurements are also shown in figure \ref{fig:anout}.\\     
It is important to have a threshold distribution as uniform as possible. On one hand the threshold must be set well above the amplifier noise in order not to flood the system with random hits. On the other hand the position resolution depends almost linearly on the threshold (see \cite{TDR}). The width of a gaussion fit to the untrimmed threshold distribution of a ROC is around 300 electrons as shown in figure \ref{fig:trim} (a). Adjusting the trim bits is one of the most important steps in the module characterization process \cite{andrei}. 
\begin{figure}[h]
\begin{center}
\includegraphics[width=7.5cm]{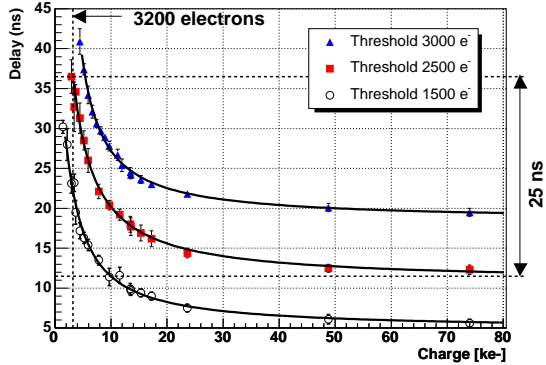}
\end{center}
\caption{Timewalk for 3 different thresholds. The offsets of the 3 curves are arbitrary for better readability}
\label{fig:timewalk}
\end{figure}
After trimming, the threshold dispersion can be improve considerably to typically 80 electrons as shown in figure \ref{fig:trim} (b). The intrinsic noise of the preamp/shaper has been measured to be $\lesssim 190$ electrons \cite{andrei}.\\
The threshold described so far is not directly relevant. The time between a particle crossing the sensor and the comparator going above threshold depends on the created ionization charge. Therefore it might happen that for low pulses the hit signal is delayed to the next bunch crossing and hence is lost during trigger validation. This timewalk can be measured by delaying the injected calibration pulses with respect to the chip clock. The result of such scans is shown in figure \ref{fig:timewalk} for three different thresholds. For an absolute threshold of 2500 electrons it is shown that the effectively usable threshold (in-time threshold) is at about 3200 electrons, i.e. the time difference between a signal of 3200 electrons and very high charges above 100k electrons is below 25ns.  \\
Test structures containing the analog section of the PUC have been irradiated with gamma rays up to a total dose of 13.2 Mrad. This corresponds to the absorbed dose in 2 years at high luminosity LHC operation for the 7cm barrel layer or 1 year for the innermost layer. Figure \ref{fig:irrad} shows the outputs of the shaper and the sample-and-hold circuit in transparent mode. Curves (a) show the response for unirradiated structures. After irradiation, the analog signals become somewhat slower (b), but can be brought back close to the original waveform by reprogramming the feedback settings (c).\\

\section{Summary}
The final version of the CMS pixel readout chip has been designed in a $0.25\mu$m process. It is fully functional and measurements of the performance have been presented. The threshold can be tuned to values below 2500 electrons with a uniformity of 80 electrons. Due to timewalk the in-time threshold is somewhat higher at 3200 electrons. The readout uses an analog scheme with address levels coded into 6 levels. This has been proven to work reliably at 40 MHz as address decoding was 100\% correct.

\end{document}